\documentclass[useAMS,usenatbib]{mn2e}

\usepackage{graphicx}
\usepackage{txfonts}

\title[Regular chains of star formation complexes]
      {Regular chains of star formation complexes in spiral arms of NGC~628}

\author[A.~S.~Gusev and Yu.~N.~Efremov]
       {A.~S.~Gusev and Yu.~N.~Efremov \\
        Sternberg Astronomical Institute, Lomonosov Moscow State University,
        Universitetsky pr. 13, 119992 Moscow, Russia, 
        {\sf gusev@sai.msu.ru} \\
             }

\date{Accepted 2013 June 6. Received 2013 June 6; in original 
form 2013 January 26}

\begin{document}

\maketitle

\begin{abstract}
We investigate photometric properties of spiral arms and stellar 
complexes/associations inside these arms in the grand design NGC~628 (M~74) 
galaxy. We analyze {\it GALEX} ultraviolet, optical $UBVRI$, and H$\alpha$ 
surface photometry data, including those obtained with 1.5~m telescope at 
the Maidanak Observatory. In the longer arm, the large and bright stellar 
complexes are located at regular intervals along the arm, but only farther 
from the galaxy center. They are joined with the narrow lane of dust, visible 
only in the infrared bands. The usual dust lane along the stellar arm inner 
side is seen there only at distances closer to the galaxy center. It is
well expressed in CO (H$_2$) image. We have found, that the second, 
short arm hosts two dust lanes, the strong and wide at the inner 
side, and narrow and irregular along its outer edge. This outer dust lane is 
well seen in IR images only. The shorter arm contains no star complexes at all. 
Gradients of age and luminosity of stars across both arms are missing (again 
excepting the parts of arms located closer to the center), which is 
confirmed by our photometric cuts across both arms. The drastic difference in 
the morphology of the two symmetric arms (grand design type) of a galaxy 
has now been confirmed by objective measurements in the case of M~74. 
It is unclear why about two third of galaxies with beaded arms host these 
''beads'' (star complexes) in one arm only.
\end{abstract}


\section{Introduction}

The young stars and clusters, as well as H\,{\sc ii} regions are known to 
be distributed along the spiral arms of grand design galaxies 
non-uniformly; rather often they form groupings with sizes of about 
0.3-0.7~kpc, rarely a little more. Older objects (such as Cepheids, 
most of which are ten times older than O stars) are usually also 
gathered in the same groups, called star complexes \citep{efremov1979}. 
These complexes are the greatest coherent groupings of stars and clusters, 
which are connected by unity of an origin from the same H\,{\sc i}/H$_2$ 
supercloud \citep{efremov1989,efremov1995,elmegreen1994,elmegreen2009,
odekon2008,marcos2009}.

In irregular galaxies, there is a continuous sequence of star groupings 
with increasing age (of the oldest stars) and size, starting from clusters 
to associations to complexes \citep{efremov1998}; in flocculent 
galaxies the largest complexes transform to short spiral segments 
\citep{elmegreen1996}. Within the regular spiral arms of grand design 
galaxies, star complexes sometimes are located along an arm at rather 
regular distances. This is a quite rare phenomenon, which was 
found by \citet{elmegreen1983} in 22 galaxies. They noted the 
spacing of complexes (which they called H\,{\sc ii} regions) in studied 
galaxies to be within 1--4 kpc, and each string to consist, on average, of 
five H\,{\sc ii} regions. The gravitational or magneto-gravitational 
instability developing along the arm was suggested to explain this 
regularity \citep{elmegreen1983,elmegreen1994}.

\begin{figure*}
\resizebox{0.93\hsize}{!}{\includegraphics[angle=000]{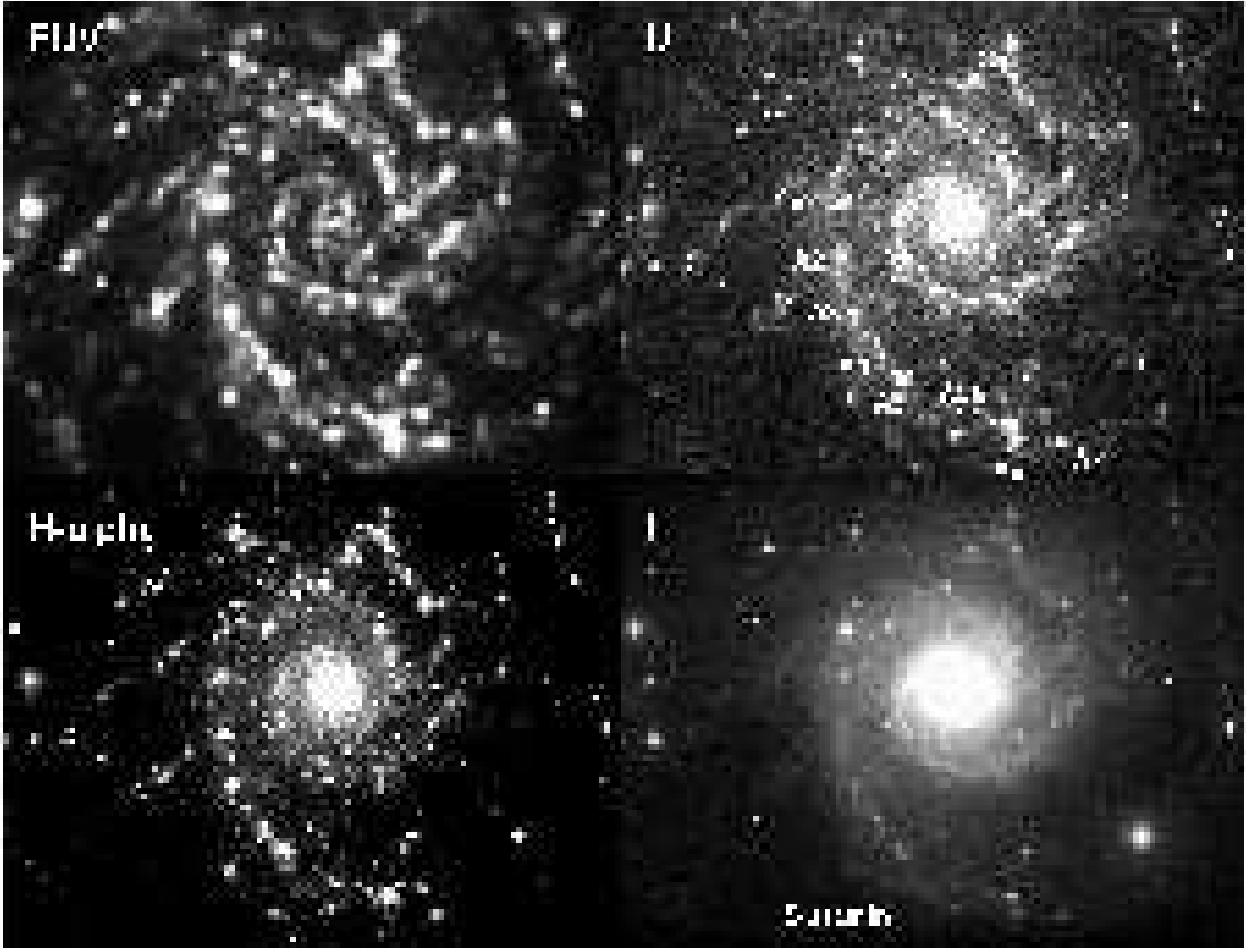}}
\caption{
Images of NGC~628 in the $FUV$ (top-left), $U$ (top-right), $I$ 
(bottom-right) passbands, and H$\alpha$ line (bottom-left). 
The $FUV$ image was taken from {\it GALEX} archive, the $U$, $I$ and 
H$\alpha$ images were obtained on the Maidanak Observatory with the 1.5~m 
telescope. Positions and ID numbers of star formation complexes from the 
list of \citet{elmegreen1983} are indicated in the $U$ image. North is 
upward and east is to the left.
{\bf High resolution jpeg image is available in the source format.}}
\label{figure:fig1}
\end{figure*}

The list of \citet{elmegreen1983} is still the only published result of 
systematic searches for string of complexes in spiral arms. Note that in 
15 of their 22 galaxies the regular strings of complexes are seen in one 
arm only. We believe, to study properties of such galaxies is a most 
perpective way to understand the nature of regular chains formation.

\begin{table}
\caption[]{\label{table:param}
Basic parameters of NGC~628.
}
\begin{center}
\begin{tabular}{ll} \hline \hline
Parameter                                & Value \\
\hline
Type                                     & Sc \\
Total apparent $B$ magnitude ($B_t$)     & 9.70 mag \\
Absolute $B$ magnitude ($M_B$)$^a$       & -20.72 mag \\
Inclination ($i$)                        & $7\degr$ \\
Position angle (P.A.)                    & $25\degr$ \\
Heliocentric radial velocity ($v$)       & 659 km/s \\
Apparent corrected radius ($R_{25}$)$^b$ & 5.23 arcmin \\
Apparent corrected radius ($R_{25}$)$^b$ & 10.96 kpc \\
Distance ($d$)                           & 7.2 Mpc \\
Galactic absorption ($A(B)_{Gal}$)       & 0.30 mag \\
Distance modulus ($m-M$)                 & 29.29 mag \\
\hline
\end{tabular}\\
\end{center}
\begin{flushleft}
$^a$ Absolute magnitude of a galaxy corrected for Galactic extinction and
inclination effect. \\
$^b$ Isophotal radius (25 mag\,arcsec$^{-2}$ in the $B$-band) corrected for 
Galactic extinction and absorption due to the inclination of NGC~628.
\end{flushleft}
\end{table}

One of such galaxies is M~31, as it is seen in the {\it GALEX} UV 
images. \citet{efremov2009,efremov2010} noticed a regular string of star 
complexes there, located in the north-western arm; it has about the same 
size -- 0.6~kpc with spacing of 1.1~kpc. Within the same arm segment, 
a regular magnetic field with a wavelength of 2.3~kpc was found earlier by 
\citet{beck1989}. This wavelength is twice as large as the spacing between 
complexes and suggests that they were formed as a result of the 
magneto-gravitational instability developed along the arm. In this 
north-western arm segment, star complexes are located inside the gas-dust 
lane, whilst in the south-western arm of M~31 the gas-dust lane is upstream 
of the bright and uniform stellar arm. Earlier, evidence for the age gradient 
was found in the south-western arm. All these are signatures of a spiral 
shock, which may be associated with an unusually large (for M~31) pitch angle 
of this south-western arm segment. Such a shock may prevent the formation 
of a regular magnetic field, which might explain the absence of star 
complexes there. The  fragmentation of the outer H\,{\sc i} arms of our 
Galaxy into regularly spaced superclouds  was also described 
\citep{efremov2011}; a magnetic field parallel to the Galactic planes is 
known there but no details are available yet.

Anticorrelation between shock wave signatures and the presence of star 
complexes is observed in spiral arms of a few other galaxies. Regular 
chains of star complexes/superclouds in spiral arms are rare, which may 
imply that a specific mechanism is involved in their formation; the most 
probable is the Parker-Jeans instability. Unfortunately, there are no 
sufficiently detailed magnetic field data (or no such data at all) to 
compare them with the arm structures.

Amongst nearby galaxies, the most evident (after M~31) case of the 
above-mentioned anticorrelation is NGC~628 (M~74) galaxy. It is a 
nearby spiral galaxy viewed almost face-on (Fig.~\ref{figure:fig1}). It is 
also an excellent example of a galaxy with regular strings of complexes 
which are seen in one arm only. It hosts two regular arms, the longer 
one of which includes a number of star complexes with similar spacing. 
In this longer arm \citet{elmegreen1983} found seven 
bright complexes (H\,{\sc ii} regions) plus a fainter one with a 
characteristic separation of 1.6-1.7~kpc (Fig.~\ref{figure:fig1}).
Another arm does not contain evident complexes, but instead the strong 
dust lane is seen along the arm inner side (and the fainter one along the 
outer side).

\begin{figure}
\vspace{5.5mm}
\resizebox{0.95\hsize}{!}{\includegraphics[angle=000]{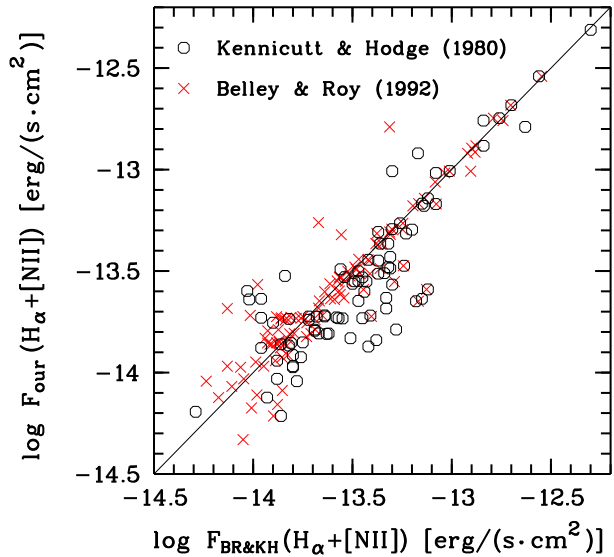}}
\caption{Comparison between our data (ordinate axis) and data of 
\citet{belley1992,kennicutt1980} (abscissa axis) of spectrophotometric
H$\alpha$+[N\,{\sc ii}] fluxes for H\,{\sc ii} regions in NGC~628.
}
\label{figure:fig3}
\end{figure}

Here, we consider a regularity of distribution in spacing and photometric 
properties of star formation regions (SFRs) in well-known grand 
design galaxy NGC~628, based on our own observations in the $U$, $B$, $V$, 
$R$, $I$ passbands, and H$\alpha$ line, as well as {\it GALEX} far- and 
near-ultraviolet ($FUV$ and $NUV$) data. We use the term ''star 
formation regions'', which includes young star complexes, young clusters, 
OB-associations, H\,{\sc ii} regions, i.e. all young stellar groups 
independently from their sizes and luminosities.

The fundamental parameters of NGC~628 are presented in 
Table~\ref{table:param}. We take the distance to NGC~628, obtained in 
\citet{sharina1996,vandyk2006}. We used the position angle and the 
inclination of the galactic disk, derived by \citet{sakhibov2004}, based 
on a Fourier analysis of the spatial distribution of the radial velocities 
of the gas in the disk. Other parameters were taken from the {\sc leda} data 
base\footnote{http://leda.univ-lyon1.fr/} \citep{paturel2003}. We adopt the 
Hubble constant H$_0 = 75$ km\,s$^{-1}$Mpc$^{-1}$ in the paper. With the 
assumed distance to NGC~628, we estimate a linear scale of 34.9~pc/arcsec.

\section{Observations and reduction}

The results of $UBVRI$ photometry for NGC~628 have already been 
published in \citet{bruevich2007}. Here we present H$\alpha$ 
spectrophotometric observations and data reduction for the galaxy, that 
have not been published earlier. Just a brief compilation is given for our 
older observations, as well as for {\it GALEX} ultraviolet data.

\subsection{Observations}

The photometric observations were obtained in September 2002 with the 1.5~m 
telescope of the Maidanak Observatory (Institute of Astronomy of the Academy 
of Sciences of Uzbekistan) using a SITe-2000 CCD array. The focal length of 
the telescope is 12~m. Detailed description of the telescope and the CCD 
camera can be found in \citet{artamonov2010}. With broadband $U$, $B$, $V$, 
$R$, and $I$ filters, the CCD array realizes a photometric system close to 
the standard Johnson--Cousins $UBVRI$ system. The camera is cooled with 
liquid nitrogen. The size of the array is $2000\times800$ pixels. It 
provides the $8.9\times3.6$ arcmin$^2$ field of view with the image scale 
of 0.267 arcsec/pixel. The seeing during the observations was 0.7--1.1 
arcsec.

Since angular size of NGC~628 was larger than the field of view, we 
obtained separate images for the northern and southern parts of the galaxy.

Spectrophotometric H$\alpha$ observations of NGC~628 were made on 2006 
September 26 with the 1.5-m telescope of the Mt.~Maidanak Observatory with 
the SI-4000 CCD camera. The chip size, $4096\times4096$ pixels, provides a 
field of view of $18.1\times18.1$ arcmin$^2$, with an image scale of 0.267 
arcsec/pixel. The total exposure time was 1200~s, the seeing was 1.0 arcsec.

The wide-band interference H$\alpha$ filter ($\lambda_{eff}$ = 6569\AA, 
FWHM = 44\AA) was used for the observations. Indicated filter parameters 
provides H$\alpha$+[N\,{\sc ii}]$\lambda$6548+$\lambda$6584 imaging for 
nearby galaxy NGC~628.

Ultraviolet {\it GALEX} $FUV$ and $NUV$ reducted FITS-images of NGC~628 were 
downloaded from Barbara A. Miculski archive for space telescopes 
(galex.stsci.edu; source GI3\_050001\_NGC628). The observations were made in 
October-November 2007. The total exposure time was 12004.6~s for every band. 
The description of the {\it Galaxy Evolution Explorer} mission and basic 
parameters of passbands were presented in \citet{morrissey2005}. We only note 
here, that an effective wavelength $\lambda_{eff} = 1528$\AA\, for the $FUV$ 
band and 2271\AA\, for the $NUV$ band, and image resolution is equal to 
4.5 arcsec for $FUV$ and 6.0 arcsec for $NUV$.

\subsection{Data reduction}

The reduction of the photometric and spectrophotometric data was carried
out using standard techniques, with the European Southern Observatory
Munich Image Data Analysis System\footnote
{http://www.eso.org/sci/software/esomidas/} ({\sc eso-midas}) 
\citep{banse1983,grosbol1990}. The main image reduction
stages for spectrophotometric data were as follows: correction for bias 
and flat field; removal of cosmic-ray traces; determining the sky 
background, then subtracting it from each image frame; aligning and 
adding the images; absolute calibration. The main photometric $UBVRI$ image 
reduction stages were described in detail in \citet{bruevich2007}.

\begin{figure}
\vspace{1.0mm}
\resizebox{1.00\hsize}{!}{\includegraphics[angle=000]{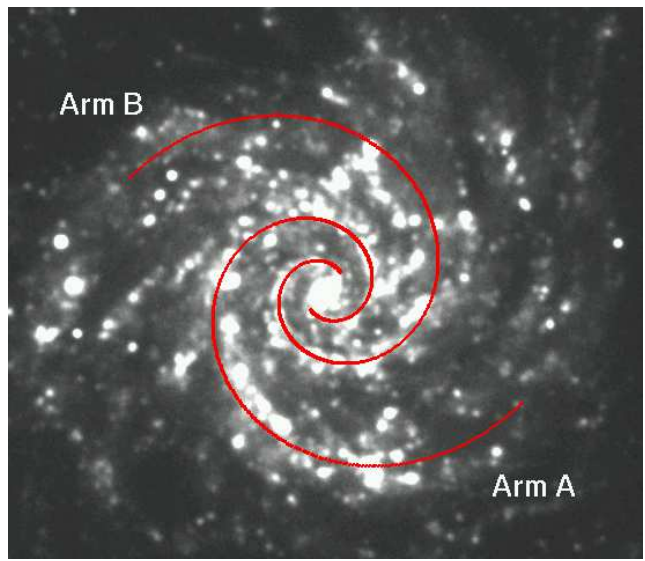}}
\caption{$NUV$ image of NGC~628 with overlaid logarithmic spirals. 
The size of the image is $9.7\times8.5$ arcmin$^2$. North is upward and 
east is to the left.
}
\label{figure:nfig2}
\end{figure}

Calibration of H$\alpha$ fluxes the was made based on the results of 
spectrophotometric observations of \citet{belley1992,kennicutt1980}. 
We calculated fluxes of H\,{\sc ii} regions using our images with apertures 
of \citet{belley1992,kennicutt1980} and found coefficient of reduction from 
our instrumental H$\alpha$ flux to the flux in physical units 
(erg\,s$^{-1}$\,cm$^{-2}$):
\begin{eqnarray}
F = (3.70\pm0.60)\times10^{-18} {\rm erg \,s^{-1}\,cm^{-2}/ADU} \nonumber
\end{eqnarray}
by data of \cite{belley1992}, and
\begin{eqnarray}
F = (4.48\pm1.48)\times10^{-18} {\rm erg \,s^{-1}\,cm^{-2}/ADU} \nonumber
\end{eqnarray}
by data of \cite{kennicutt1980}. The value
$3.70\times10^{-18}$ for the reduction coefficient was adopted.

Fig.~\ref{figure:fig3} compares H$\alpha$+[N\,{\sc ii}] fluxes derived in 
the present paper with the fluxes observed by 
\cite{belley1992,kennicutt1980}. The data are not corrected for 
interstellar reddening.

For the absolute calibration of {\it GALEX} data, we used zero point levels 
for $FUV$ and $NUV$ images \citep{morrissey2005}.

We corrected all data for Galactic absorption; these values are indicated 
by an ''0'' subscript. We used the resulting ratio of the extinction in the 
{\it GALEX} bands to the color excess $E(B-V)$ are $A_{FUV}/E(B-V)=8.24$ 
and $A_{NUV}/E(B-V)=8.2$ \citep{wyder2007}.

\section{Spiral arms}

\subsection{Parameters of spiral arms}

We investigate properties of spiral arms by curve fitting of two principal 
arms, which are clearly outlined in blue ($FUV$, $NUV$, $U$, $B$) passbands 
and H$\alpha$ line (Fig.~\ref{figure:fig1}). The arms are defined 
by by-eye selection of pixels in the part of these images that is 
within the spiral arms. Pixels in this region are then fitted with a 
logarithmic spiral using linear least-squares.

A logarithmic spiral with a pitch angle $\mu$ can be described as
\begin{equation}
r = r_0e^{k(\theta-\theta_0)},
\label{equation:spiral}
\end{equation}
where $k = \tan \mu$. We adopt two constraints for the fitting: the pitch 
angle is constant along the arm, and both spiral arms are symmetric. 

We do not transform images of NGC~628 to the face-on position, 
$i=0\degr$, because the position of any point in the galaxy is changed 
by value much less than the characteristic width of the spiral arm, if 
we use the adopted inclination $i=7\degr$. Difference between the uncorrected 
position and the position corrected for the inclination position angle 
is less than $0.5\degr$ and the galactocentric distance 
$\Delta r/r < 0.01$ for any point in the galaxy.

Spiral arm with a regular string of complexes found by \citet{elmegreen1983} 
was named Arm~A, and the opposite arm was named Arm~B 
(Fig.~\ref{figure:nfig2}). Arm~A is known as Arm~2 in 
\citet{kennicutt1976,cornett1994} or South arm in \citet{rosales2011}.

As a result of the fitting, we obtained coefficients in 
Eq.~(\ref{equation:spiral}) $r_0 = 21.95$ arcsec (765~pc) and $k = 0.280$, 
that corresponds to a pitch angle $\mu = 15.7\degr$, for the arms. Zero 
point angle $\theta_0 = -40\degr$ for Arm~A and $140\degr$ for Arm~B (where 
$\theta$ is counted from north toward east). The resulting spiral arms 
are shown in Fig.~\ref{figure:nfig2}. Obtained pitch angle, 
$\mu = 15.7\degr$, is close to the result of \citet{kennicutt1976} who 
fitted every arm separately and obtained values $13.8\degr$ for Arm~A and 
$11.2\degr$ for Arm~B.

\subsection{Along-arm photometry}

\begin{figure*}
\vspace{6.2mm}
\resizebox{0.85\hsize}{!}{\includegraphics[angle=000]{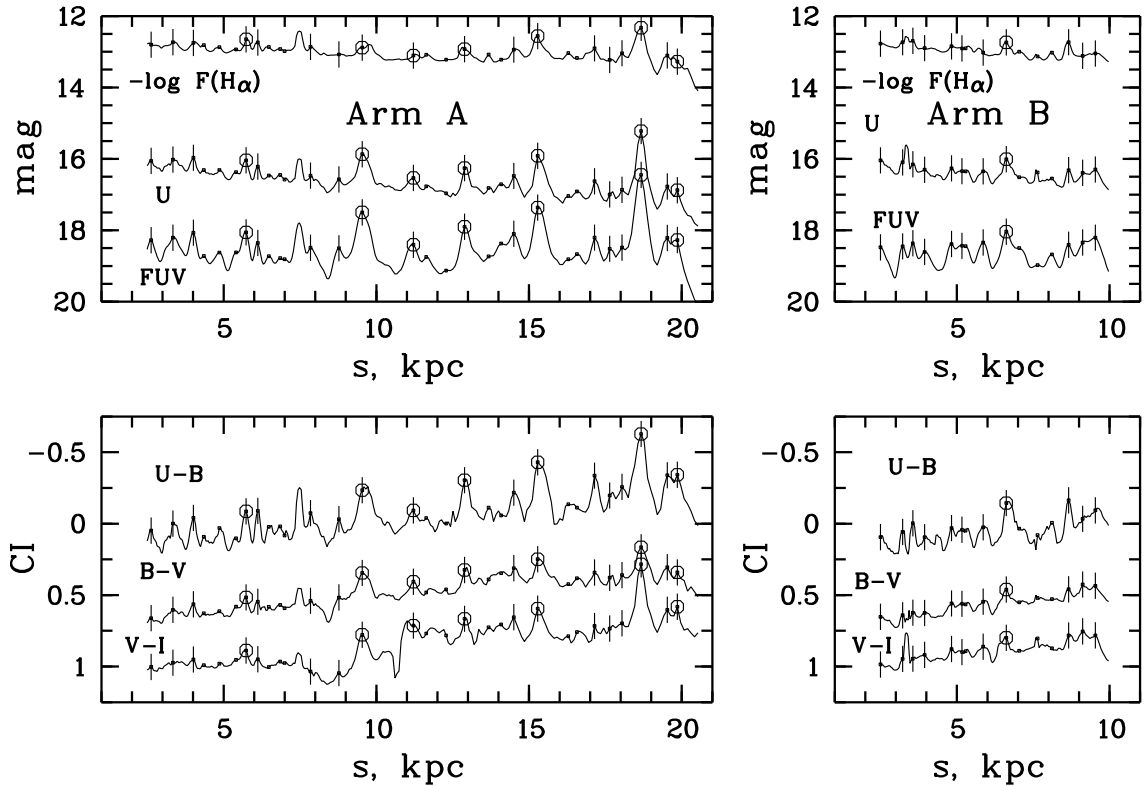}}
\caption{Photometric (top panels) and colour indices (bottom panels) profiles 
along Arm~A (left panels) and Arm~B (right panels). The units on the ordinate 
axis are corrected for Galactic absorption magnitudes, logarithm of 
H$\alpha$ flux, and colour indices within the aperture. Positions of 
local maxima of brightness (dots), star formation regions 
(vertical bars) and bright complexes (circles) are indicated.
}
\label{figure:nfig3}
\end{figure*}

To study variation of brightness along a spiral arm, we obtained 
photometric profiles along Arms~A and B. The elliptical aperture 
($40\times6$ arcsec$^2$) with a minor axis along a spiral arm (i.e. a 
difference between P.A. of major axis and P.A. of the centre of aperture is 
a pitch angle), and a step of $1\degr$ by P.A. was  used. Obtained photometric 
and colour indices profiles along the arms are presented in 
Fig.~\ref{figure:nfig3}. For a logarithmic spiral in the form of 
Eq.~(\ref{equation:spiral}), the longitudinal displacement along the spiral, 
denoted as $s$, is
\begin{equation}
s = (\sin \mu)^{-1}r_0(e^{k(\theta-\theta_0)}-1).
\label{equation:sp_long}
\end{equation}

Star formation regions, visible in $FUV$, are absent in the innermost 
part of the spiral arms (Fig.~\ref{figure:fig1}). SFRs nearest to the 
galactic centre are located at distance $r\approx42 - 43$ arcsec in 
both arms. Therefore, we do not include the inner part of the spiral 
arms with a galactocentric distance $r < 41.45$ arcsec (1.45 kpc) or an 
angular sector $\theta-\theta_0 < 130\degr$ in our research. Distorted outer 
part of Arm~B was also excluded from the examination.

\begin{figure*}
\vspace{4.0mm}
\resizebox{0.85\hsize}{!}{\includegraphics[angle=000]{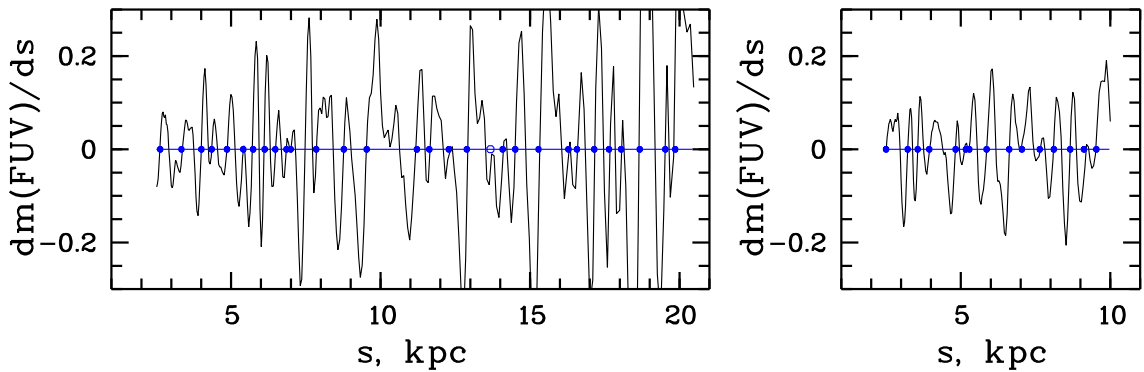}}
\caption{
Function $dm(FUV)/ds$ along Arm~A (left panel) and Arm~B (right panel).
The circles indicate local maxima of brightness. See the text for more 
explanation.
}
\label{figure:nfig4b}
\end{figure*}

Using the profile in the $FUV$ band, we have found local maxima of brightness 
on the profile. The local maxima of brightness were determined as lower 
extrema of the function $m_{FUV}(s)$. To locate them, we looked for 
points, where the first derivative of the function, $dm(FUV)/ds = 0$ and 
$d^2m(FUV)/ds^2 > 0$ on the profile (Fig.~\ref{figure:nfig4b}). Positions 
of these points also correspond to relatively bluer parts of spiral arms 
(Fig.~\ref{figure:nfig3}).

\begin{figure*}
\resizebox{0.95\hsize}{!}{\includegraphics[angle=000]{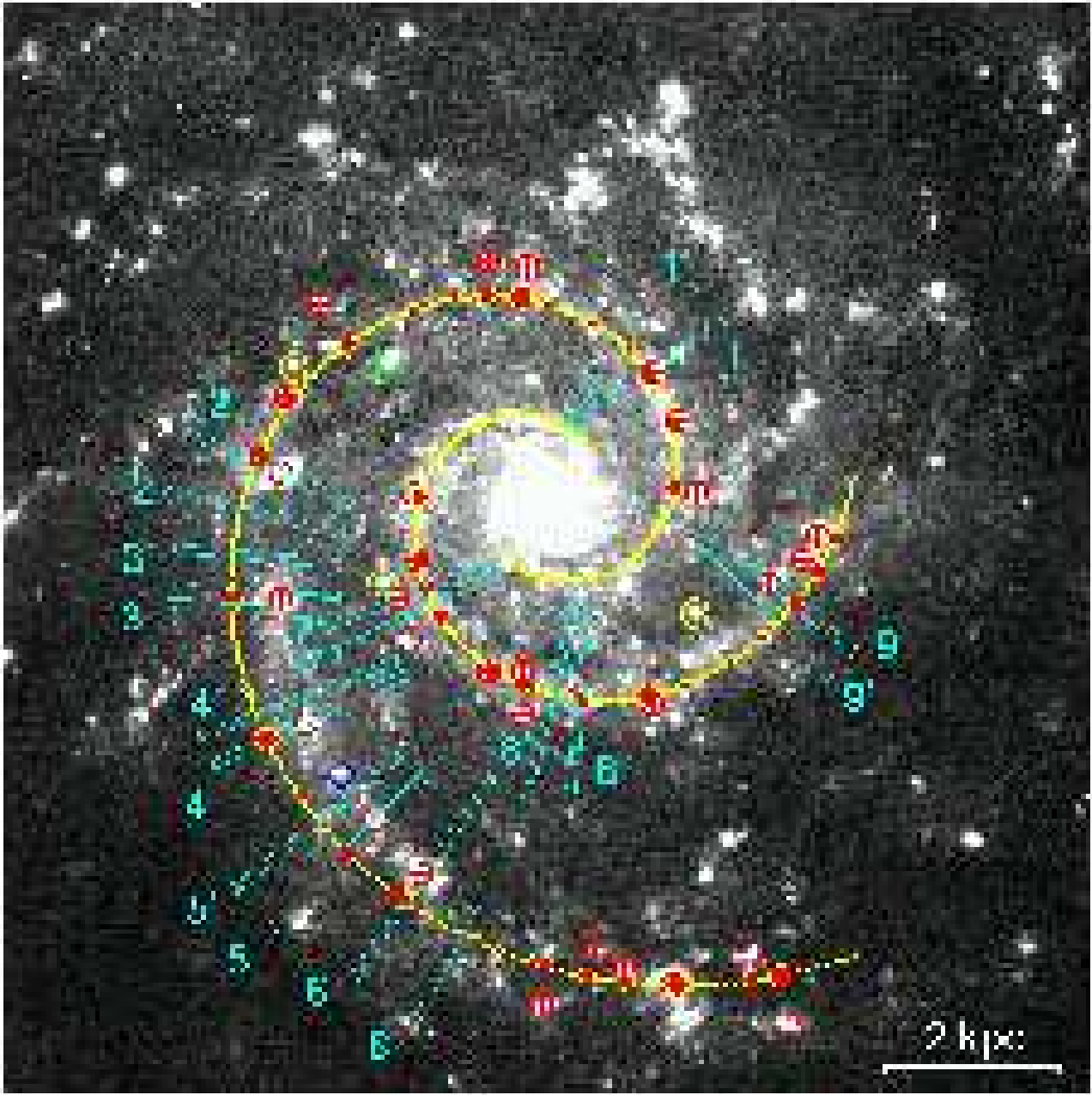}}
\caption{Image of NGC~628 in the $U$ passband. Spiral arms are shown. The 
dark (red) large, middle and small circles on the arm's curves correspond 
to the projective positions of the bright complexes, star formation 
regions and the local maxima of brightness, respectively. 
The dark (red) crosses in circles show positions of the selected SFRs 
and complexes. The positions of across-arm profiles 1,~1'--9,~9' are 
indicated. The grey (green) square shows the interarm complex. The 
white (yellow) circles show the field stars, which have been removed 
from images before the along-arm photometry. The dark (blue) diamond 
shows the position of the additional association. The size of the image is 
$6.0\times6.0$ arcmin$^2$. North is upward and east is to the left. See the 
text for more explanation. (A colour version of this figure is available in 
the online version.)
{\bf High resolution jpeg image is available in the source format.}}
\label{figure:nfig5a}
\end{figure*}

We made only one exception when selecting local maxima of brightness. 
The additional maximum with the longitudinal displacement $s = 13.7$~kpc in 
Arm~A was included in the list of local maxima. The function $dm(FUV)/ds$ 
does not attain 0 at this $s$ (open circle in Fig.~\ref{figure:nfig4b}), 
however, obvious local maximum of brightness at $s = 13.7$~kpc is 
observed on the profiles in the $U$ band and H$\alpha$ line 
(Fig.~\ref{figure:nfig3}). The object which is responsible for this local 
maximum of brightness is a compact association, it is seen very well on the 
galactic images in the ultraviolet, $U$ and H$\alpha$ line 
(Figs.~\ref{figure:fig1}, \ref{figure:nfig2}, \ref{figure:nfig5a}). It is 
shown by diamond in Fig.~\ref{figure:nfig5a}.

The maximum of brightness at the longitudinal displacement $s = 7.5$~kpc 
in Arm~A is a result of input from the bright interarm complex
(Figs.~\ref{figure:nfig3}, \ref{figure:nfig4b}, \ref{figure:nfig5a}); it
was missed.

In a few cases, a shift of the maximum of brightness in different 
profiles along the longitudinal displacement $s$ by 1 pixel is observed for 
the same objects; we adopt the position of the maximum on the $FUV$ profile 
in such cases (see Figs.~\ref{figure:nfig3}, \ref{figure:nfig4b}). 

Objects responsible for local maxima of brightness can be 
complexes, associations, star clusters and their groups. Strong maximum 
of the profile does not necessarily indicate the presence of a bright 
complex here. There may be somewhat independent associations or clusters, 
located at the same longitudinal displacement $s$ along the arm. In addition, 
the radiation from bright SFRs can inhibit the radiation 
from fainter SFRs closely spaced along the arm. This is especially 
critical for profiles in the $FUV$. Therefore, to find and select bright 
star formation regions and complexes, we measured magnitudes of 
all objects in the arms of NGC~628 responsible for the local maxima of 
brightness on the profiles.

The photometry was made in round apertures, and the light from the 
surrounding background was subtracted from the light coming from the area, 
occupied by the star formation region. The technique of star 
formation regions photometry is described in more detail in 
\citet{gusev2003,bruevich2007}.

As a result, we selected 30 star formation regions having a total magnitude 
$FUV_0 < 19.7$~mag (Fig.~\ref{figure:nfig5a}). Following 
\citet{elmegreen1983}, we divided the objects into bright complexes and 
fainter SFRs. Eight complexes brighter than 18.1~mag in $FUV$ were selected 
as ''bright complexes'', other 22 objects were named ''star formation 
regions''. Unfortunately, we could not use ''true'' brightnesses, 
corrected for interstellar absorption, because extinction data, 
obtained from spectroscopic and spectrophotomtric observations, are known 
not for all objects. We plan to study photometric properties, 
chemical abundances, sizes and estimation of ages of brightest star 
formation regions in Arms~A and B in the further paper.

The choice of lower limit of magnitude for bright complexes and star 
formation regions was subjective. We cut the list of bright complexes 
on the seventh brightest complex in Arm~A; it coincides with the 
bright H\,{\sc ii} regions list of \citet{elmegreen1983} with one 
exception (Figs.~\ref{figure:fig1}, \ref{figure:nfig5a}). 
Among the regions fainter 19.7~mag in $FUV$, we found 
a large number of diffuse objects without a strong H$\alpha$ emission. 
Obviously, these regions are not complexes as determined in 
\citet{efremov1979,efremov1995}. We will show below, that the variation 
of the limits of brightness does not affect our conclusion on principle.

\begin{figure*}
\vspace{6.5mm}
\resizebox{0.90\hsize}{!}{\includegraphics[angle=000]{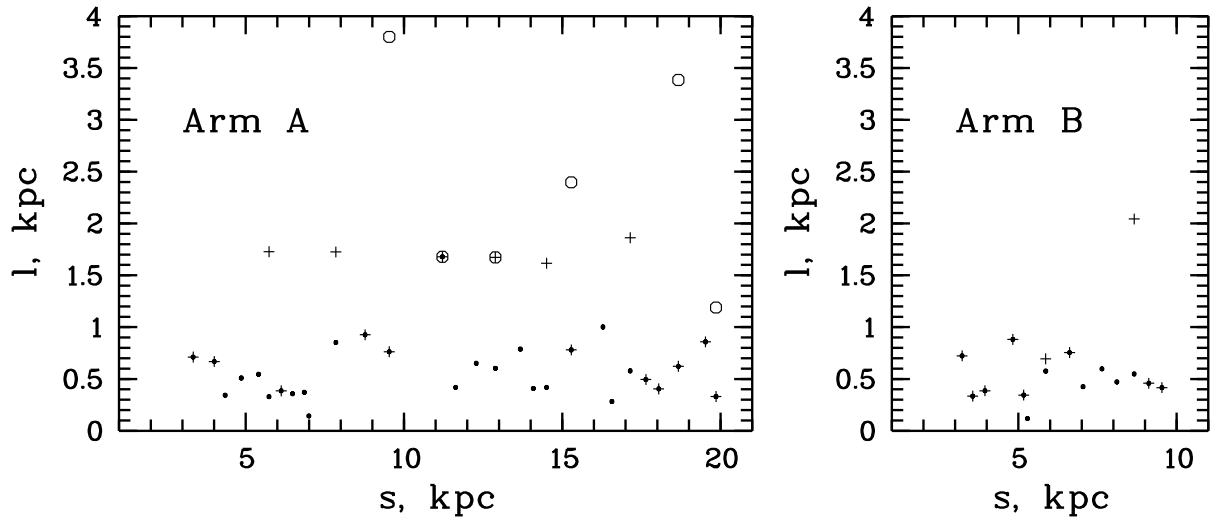}}
\caption{Separations $l$ between adjacent local maxima of brightness (dots), 
star formation regions (crosses), and bright complexes (circles) along 
Arm~A (left panel) and Arm~B (right panel). See the text for more explanation.
}
\label{figure:nfig6}
\end{figure*}

\begin{figure}
\vspace{3.6mm}
\resizebox{0.90\hsize}{!}{\includegraphics[angle=000]{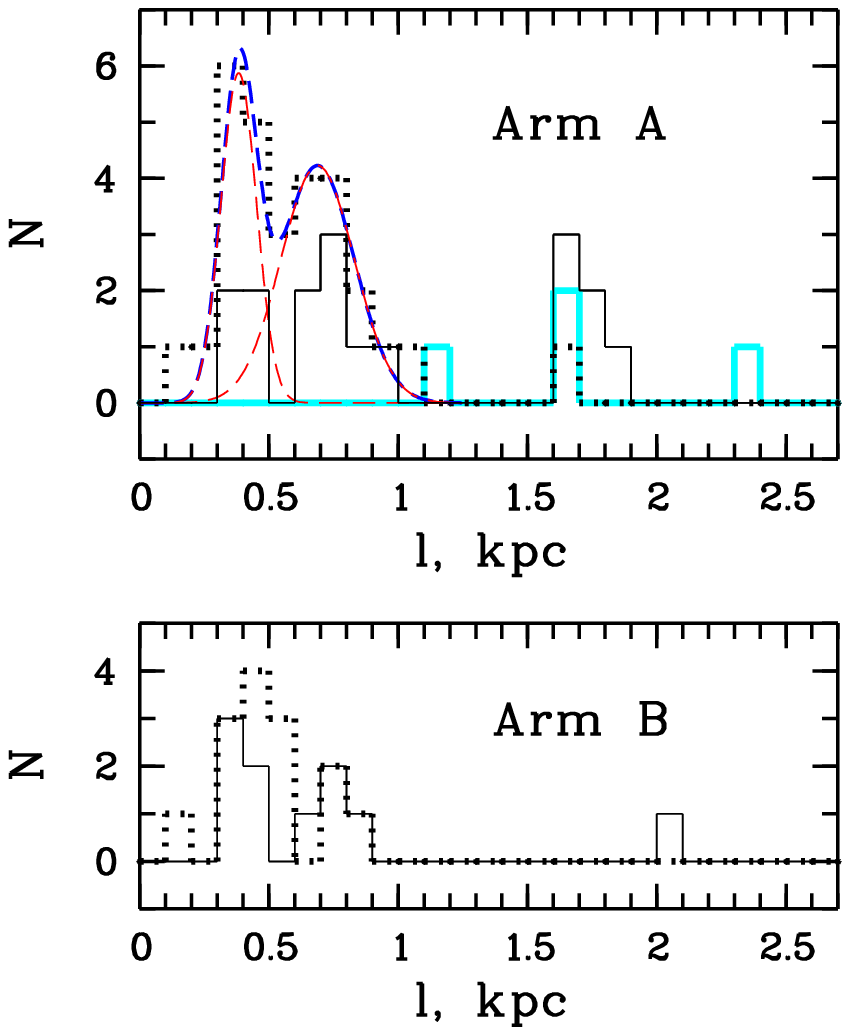}}
\caption{Number distribution histograms of the local maxima of brightness 
(thick dotted line), star formation regions (thick grey (cyan) solid 
line), and bright complexes (thin black solid line) by separation between 
adjacent objects along Arm~A (top panel) and Arm~B (bottom panel). The 
best fit Gaussians (grey (red) dashed lines) and their sum (thick dark 
(blue) dashed line) for the distribution of the local maxima of brightness 
in Arm~A are shown.
}
\label{figure:nfig7}
\end{figure}

\begin{figure}
\vspace{0.6mm}
\resizebox{0.90\hsize}{!}{\includegraphics[angle=000]{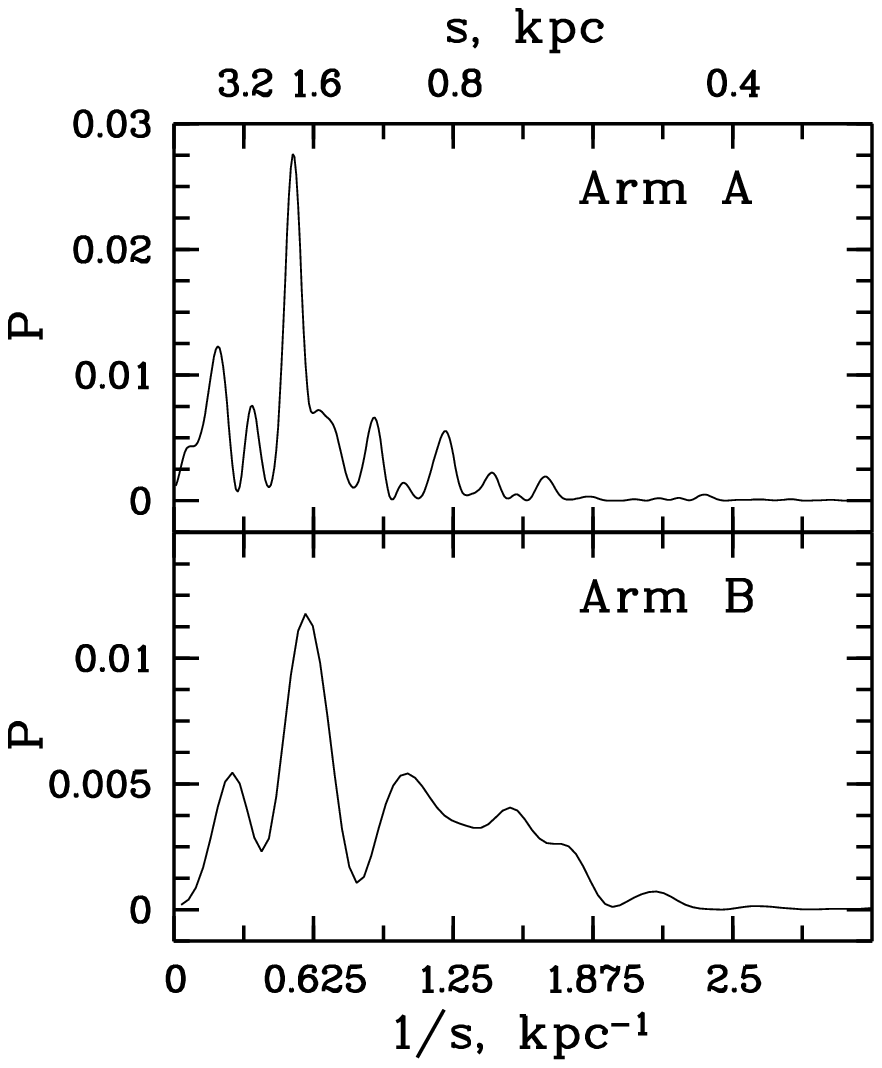}}
\caption{Power spectra of the $FUV$ profile data from Fig.~\ref{figure:nfig3} 
for Arm~A (top panel) and Arm~B (bottom panel).
}
\label{figure:nfig7ff}
\end{figure}

\begin{figure}
\vspace{0.6mm}
\resizebox{0.90\hsize}{!}{\includegraphics[angle=000]{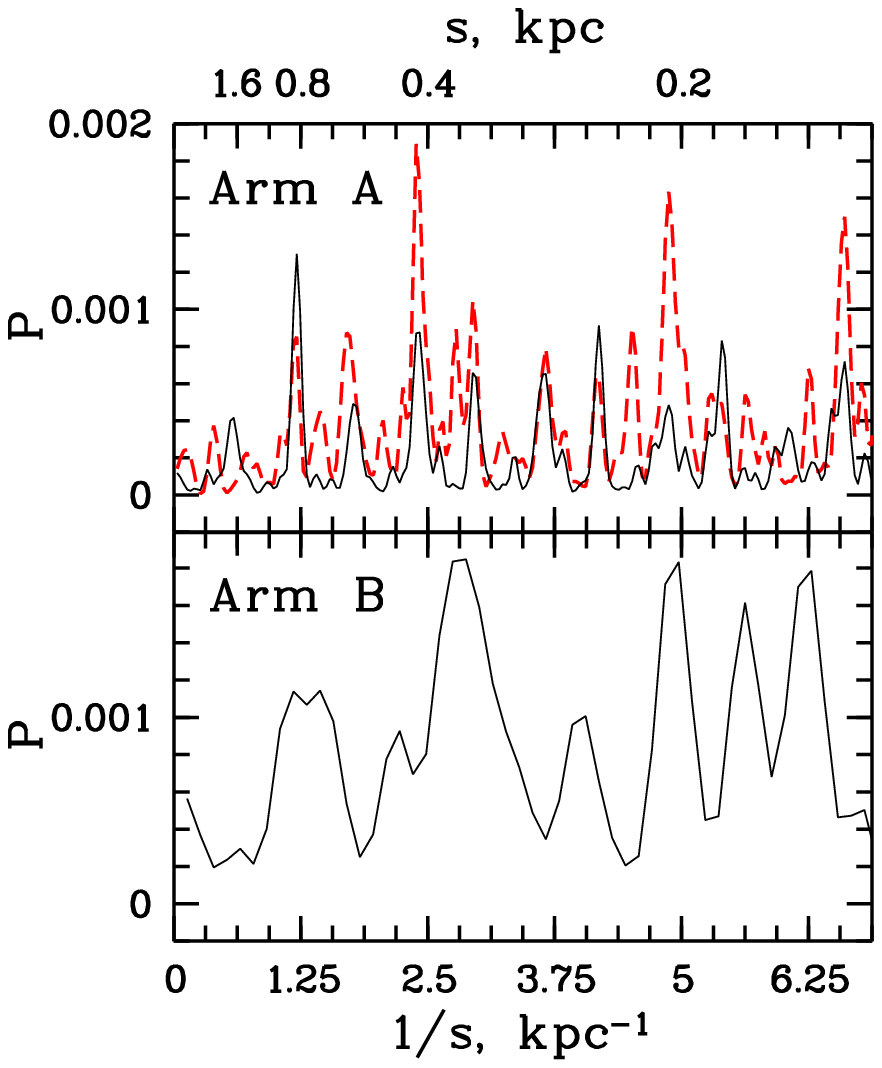}}
\caption{Power spectra of the function $D(s)$ for local maxima of 
brightness (dark (red) dashed line) and star formation regions (black solid 
line) in Arm~A (top panel), and star formation regions in Arm~B (bottom 
panel).
}
\label{figure:nfig7f}
\end{figure}

We measured separations, $l$, between adjacent bright complexes, SFRs, 
and local maxima of brightness along each arm (Fig.~\ref{figure:nfig6}). 
The set of the local maxima of brightness includes complexes and SFRs; 
the set of star formation regions includes bright complexes. Separation 
$l$ between ($n$-1)$^{st}$ and $n^{th}$ regions is defined as $s_n-s_{n-1}$, 
where $s \equiv s_n$ is defined from Eq.~(\ref{equation:sp_long}). 
Distribution of the regions by separation $l$ in Arms~A and B is presented in 
Fig.~\ref{figure:nfig7}.

As seen from the figures, the regularity in spacing of both the SFRs and 
the local maxima of brightness is observed in spiral arms of the galaxy. 
The local maxima of brightness and the star formation regions show 
a multimodal distribution of distances between adjacent objects in both 
spiral arms (Fig.~\ref{figure:nfig7}). We found a characteristic 
separation $\Lambda \approx 400$~pc for the local maxima of brightness in 
both Arms~A and B (Figs.~\ref{figure:nfig6}, \ref{figure:nfig7}). This value 
is the same in inner part ($s < 10$~kpc) of Arm~A, where bright complexes 
are absent, and in outer part of the arm, which is populated by bright 
complexes (Fig.~\ref{figure:nfig6}). The distribution of the local maxima of 
brightness by separation has two maxima, $\Lambda$ and $2\Lambda$, in 
both arms (Fig.~\ref{figure:nfig7}). There are three characteristic 
separations ($\Lambda$, $2\Lambda$ and $4\Lambda$) in Arm~A and two 
ones ($\Lambda$ and $2\Lambda$) in Arm~B are distinguished for the star 
formation regions. The characteristic separations of the bright complexes in 
Arm~A is also multiple of $\Lambda$ (Fig.~\ref{figure:nfig7}).

For additional analysis, we estimated mean and median separations 
between local maxima of brightness and the star formation regions in Arms~A 
and B for individual subsets of the objects. Results are presented in 
Table~\ref{table:mean}.

The local maxima of brightness in Arm~A at separations $l < 1$~kpc can not 
be divided into two subsets visually on the histogram. We fitted the 
distribution of the local maxima of brightness of Arm~A at $l = 0.2-1.0$~kpc 
(26 points) using two Gaussians. The parameters of obtained Gaussians (mean 
values and standard deviations) are presented in Table~\ref{table:mean}. 
The profiles of two Gaussians and their sum are also shown in 
Fig.~\ref{figure:nfig7}.

As seen from Table~\ref{table:mean}, the characteristic mean separations 
(i) are approximately equal to or multiple of $\sim0.4$~kpc, and (ii) they 
are close to each other in Arm~A and Arm~B.

Additionally, we calculated the power spectrum for the $FUV$ magnitude 
data shown in Fig.~\ref{figure:nfig3} to estimate spacing regularity 
of star formation regions in Arms~A and B (Fig.~\ref{figure:nfig7ff}). 
Obtained power spectrum for the magnitude curve in Arm~A has a basic peak 
at $s\approx2$~kpc. Secondary peaks at shorter wavelengths are observed at 
$s\approx1.6$, 1.2, and 0.8 kpc (Fig.~\ref{figure:nfig7ff}). Power spectrum 
for Arm~B has a main peak at $s\approx1.6-1.7$~kpc and a secondary peak at 
$s\approx0.9-1.0$~kpc.

To find regularity in spacing of young stellar objects on the shorter 
wavelengths, we calculated the Fourier transform of the function $D(s)$, 
where $D(s)=1$ in points of the local maximum of brightness (star 
formation region) and $D(s)=0$ in all other points. Obtained power 
spectra of the function $D(s)$ for the local maxima of brightness and 
star formation regions in Arm~A, and the star formation regions in Arm~B 
are presented in Fig.~\ref{figure:nfig7f}.

\begin{table}
\caption[]{\label{table:mean}
Characteristic separations of the local maxima of brightness and the 
star formation regions in Arms~A and B.
}
\begin{center}
\begin{tabular}{ccccc} \hline \hline
Sample & \multicolumn{2}{c}{Arm~A} & \multicolumn{2}{c}{Arm~B} \\
       & mean  & median            & mean  & median \\
       & (kpc) & (kpc)             & (kpc) & (kpc)  \\
\hline
Local maxima   & $0.38\pm0.07$ & 0.37 & $0.46\pm0.09$ & 0.43 \\
of brightness  & $0.69\pm0.14$ & 0.67 & $0.79\pm0.08$ & 0.72 \\[2mm]
Star formation & $0.40\pm0.07$ & 0.39 & $0.39\pm0.05$ & 0.35 \\
regions        & $0.76\pm0.11$ & 0.71 & $0.76\pm0.08$ & 0.72 \\
               & $1.71\pm0.08$ & 1.68 &               &      \\
\hline
\end{tabular}\\
\end{center}
\end{table}

Results of the Fourier analysis qualitatively support the estimation of 
characteristic separations of local maxima of brightness and SFRs in 
the arms based on their distribution in Fig.~\ref{figure:nfig7}. Power 
spectrum has strong peaks at $s\approx0.2$ and 0.4~kpc and weak peaks at 
$s\approx0.6$ and 0.8~kpc for the local maxima of brightness in Arm~A; 
strong peaks at $s\approx0.4$ and 0.8~kpc and weak peaks at $s\approx0.6$ 
and 1.6~kpc for the SFRs in Arm~A. Spectrum profile of SFRs in Arm~B is 
smoother than profiles of the spectra for the objects in Arm~A, 
nevertheless, peaks at $s\approx0.2$, 0.4, and 0.8~kpc stand out 
(Fig.~\ref{figure:nfig7f}).

Based on data in Table~\ref{table:mean}, and assuming that separations 
between local maxima of brightness and SFRs are approximately equal to or 
multiple of $\sim0.4$~kpc, we obtained equations of the regular 
disposition of local maxima of brightness along Arms~A and B of the 
galaxy:
\begin{equation}
s_m = 0.419s+2.800
\label{equation:sfrs_arm_a}
\end{equation}
for the local maxima of brightness in Arm~A and
\begin{equation}
s_m = 0.391s+2.421
\label{equation:sfrs_arm_b}
\end{equation}
for objects in Arm~B. Here, $s_m$ is a position of local maximum of 
brightness (star formation region) along the spiral arm, and coefficients 
$\Lambda = 0.419$, 0.391~kpc are the characteristic separations between 
local maxima of brightness and SFRs along Arm~A and Arm~B, respectively.

\begin{figure*}
\vspace{4.6mm}
\resizebox{0.85\hsize}{!}{\includegraphics[angle=000]{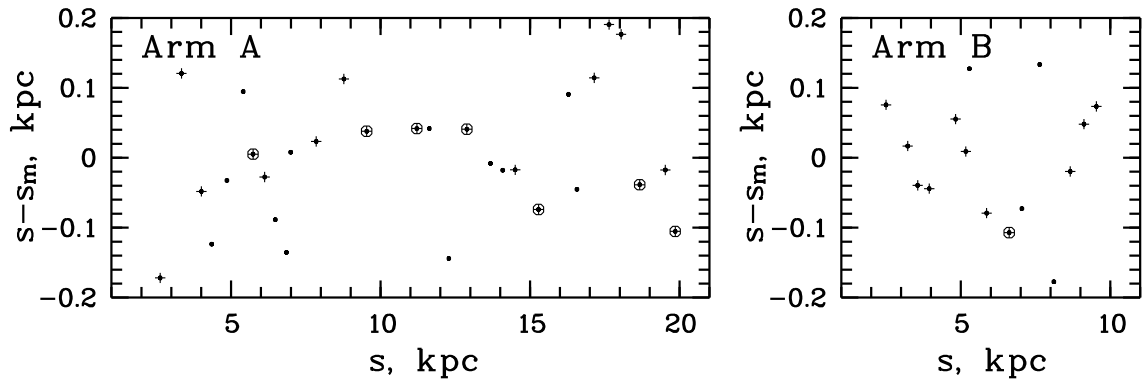}}
\caption{Deviations $s-s_m$ for the local maxima of brightness (dots), 
star formation regions (crosses), and bright complexes (circles) along 
Arm~A (left panel) and Arm~B (right panel). See the text for more 
explanation.
}
\label{figure:nfig7m}
\end{figure*}

Using values $s_m$ from Eqs.~(\ref{equation:sfrs_arm_a}) and 
(\ref{equation:sfrs_arm_b}), we calculated deviations of actual positions 
of the local maxima of brightness along the spiral arm $s$ from the 
nearest ones with positions $s_m$ (Fig.~\ref{figure:nfig7m}). The figure 
shows a good spacing regularity of star formation regions in the greater 
part of Arm~A. Average deviation $s-s_m$ for SFRs located on longitudinal 
displacement $s$ from 4 to 16~kpc is $9\pm54$~pc (Fig.~\ref{figure:nfig7m}). 
In the outer part ($s>16$~kpc), Arm~A widens, star formation regions are 
located both on the outer and the inner edge of the arm 
(Fig.~\ref{figure:nfig5a}). Disposition of star formation regions here is 
more poorly described by Eq.~(\ref{equation:sfrs_arm_a}) than for 
the central part of Arm~A. We observe that a chain of SFRs at $s=17-18$~kpc 
is displaced by $\Lambda/2 \approx 0.2$~kpc relative to other SFRs 
(Fig.~\ref{figure:nfig7m}). Deviation of positions of star formation 
regions in Arm~B from the positions $s_m$ does not exceed 100~pc; an average 
deviation $s-s_m = -11\pm62$~pc (Fig.~\ref{figure:nfig7m}).

Positions of local maxima of brightness in both spiral arms are 
slightly worse described by Eqs.~(\ref{equation:sfrs_arm_a}) and 
(\ref{equation:sfrs_arm_b}) than positions of the SFRs 
(Fig.~\ref{figure:nfig7m}). Their average deviations $s-s_m = 0\pm93$~pc 
($\pm2\Lambda/9$) in Arm~A and $0\pm88$~pc ($\pm2\Lambda/9$) in Arm~B. 
Thus, positions of most local maxima of brightness and star formation 
regions in both arms well satisfy the positions $s_m$, obtained from 
Eqs.~(\ref{equation:sfrs_arm_a}) and (\ref{equation:sfrs_arm_b}).

\begin{figure*}
\vspace{6.3mm}
\resizebox{0.95\hsize}{!}{\includegraphics[angle=000]{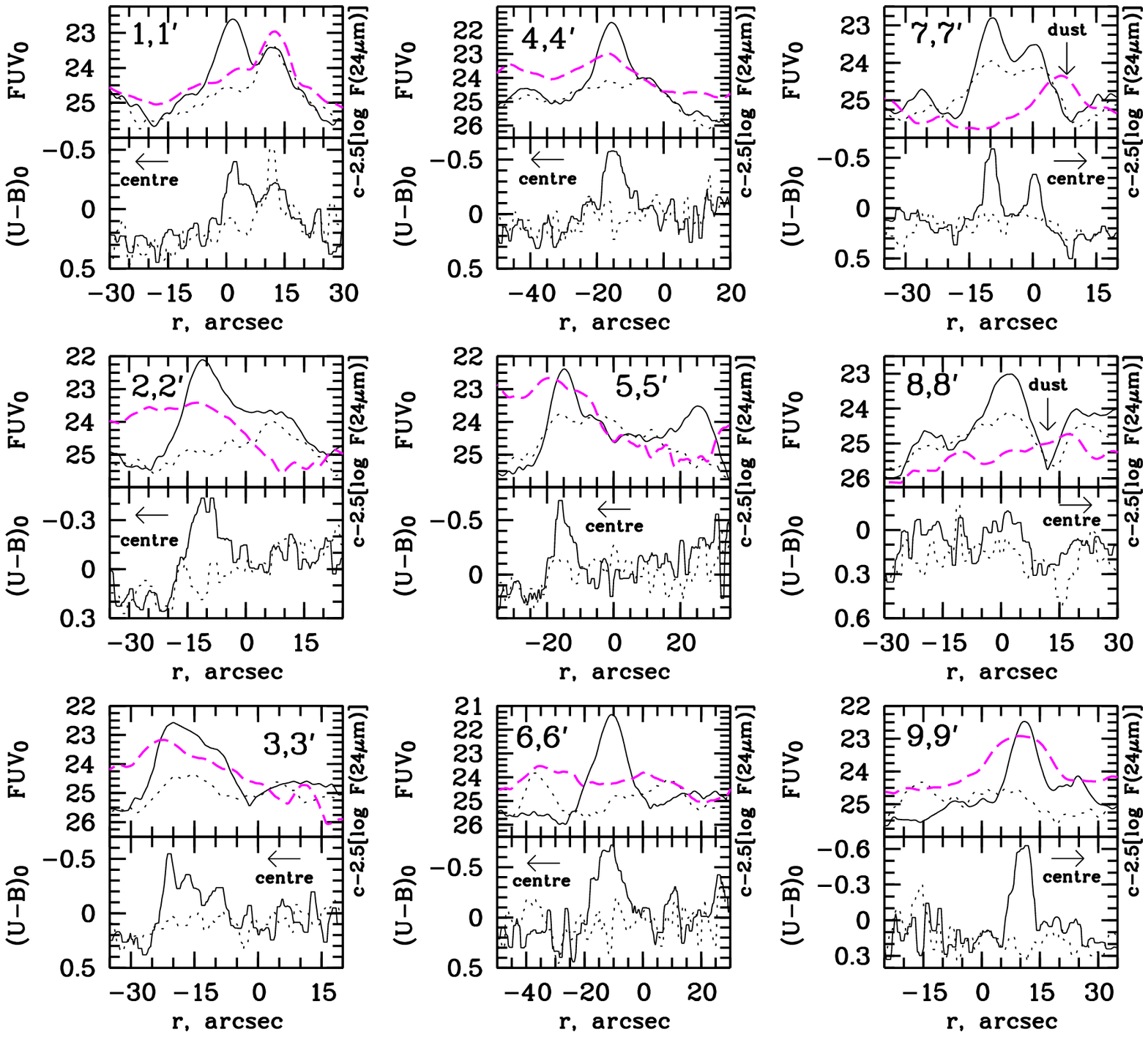}}
\caption{$FUV_0$ photometric, $(U-B)_0$ colour index (black lines), and 
$24~\mu$m (dark (magenta) lines) profiles across Arm~A (left and central 
panels) and Arm~B (right panels). Positions of profiles 1,~1'--9,~9' are 
shown in Fig.~\ref{figure:nfig5a}. Solid lines correspond to profiles 1--9, 
dotted and dashed lines correspond to profiles 1'--9'. C is an 
arbitrary constant for scaling the $24~\mu$m curves. Directions to the 
galactic centre and positions of dust lanes indicated by arrows. See the 
text for more explanation.
}
\label{figure:nfig18}
\end{figure*}

\begin{figure*}
\vspace{6.3mm}
\resizebox{0.95\hsize}{!}{\includegraphics[angle=000]{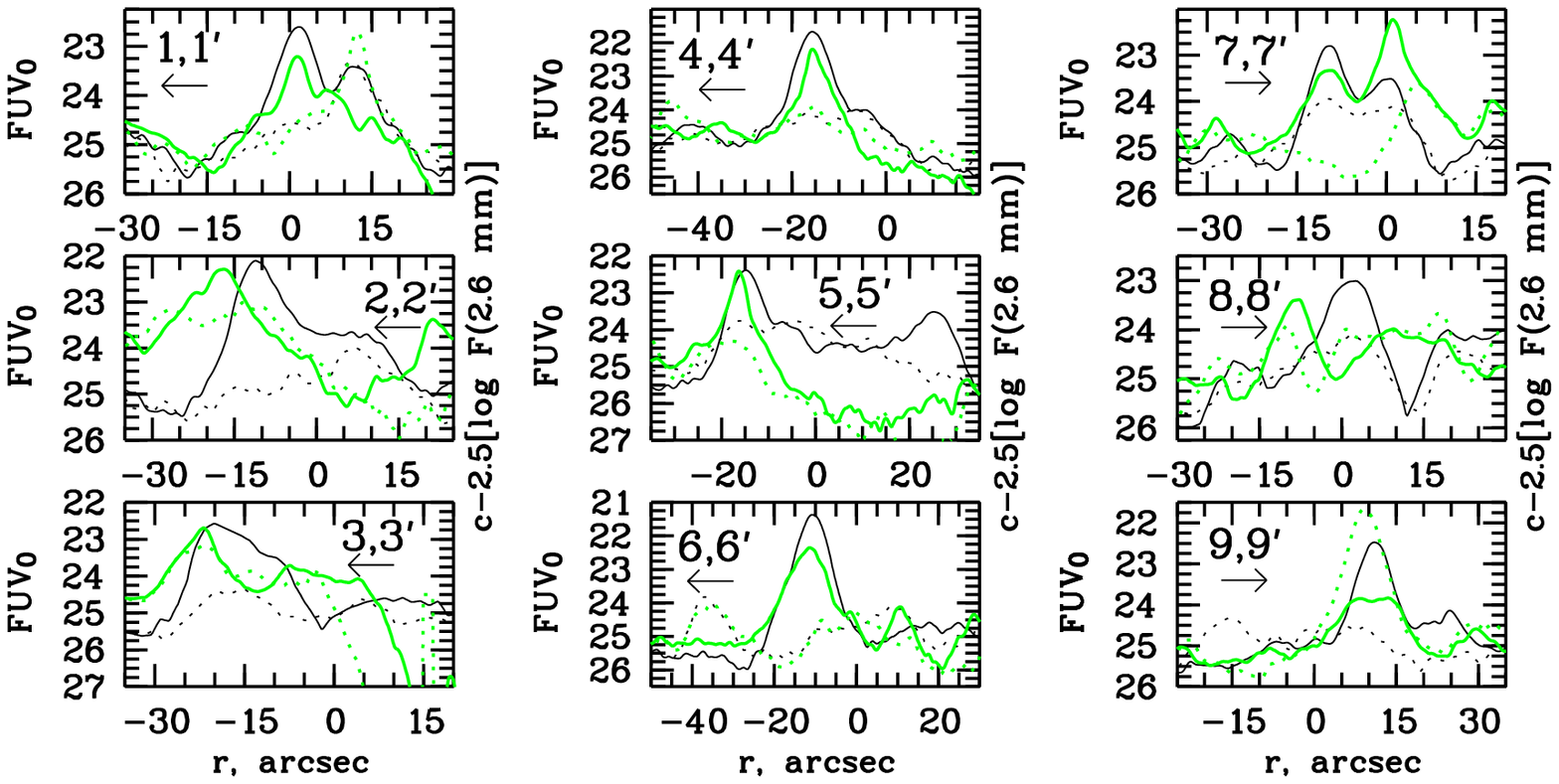}}
\caption{$FUV_0$ photometric and 2.6~mm flux (CO; grey (green) lines) 
profiles across Arm~A (left and central panels) and Arm~B (right panels). 
Solid lines correspond to profiles 1--9, dotted lines correspond to profiles 
1'--9'. C is an arbitrary constant for scaling the CO curves. Directions to 
the galactic centre indicated by arrows.
}
\label{figure:nfig18co}
\end{figure*}

Different used techniques complement each other to support presence of 
the regularity in spacing of SFRs in spiral arms of NGC~628. Power spectrum 
for the ultraviolet magnitude curve shows a periodicity at 
$s\approx1.6-2.0$~kpc ($4\Lambda-5\Lambda$) in both spiral arms 
(Fig.~\ref{figure:nfig7ff}). Power spectrum of the function $D(s)$ shows 
clear peaks at $s\approx0.4$ and 0.8 kpc ($\Lambda$ and $2\Lambda$) for the 
objects in both spiral arms (Fig.~\ref{figure:nfig7f}). Study of disposition 
of local maxima of brightness shows that the spacing regularity is observed 
throughout the parts of the spiral arms under study (one--two dozens of 
$\Lambda$; Fig.~\ref{figure:nfig7m}). We suggest that the spacing regularity 
of local maxima of brightness and SFRs exists in both spiral arms of NGC~628 
with a characteristic separation $\Lambda \approx 400$~pc. Brighter 
regions have larger characteristic separations (Fig.~\ref{figure:nfig7f} 
(top panel) illustrates it very well). The characteristic separation between 
adjacent bright complexes $l = 4\Lambda \approx 1.6-1.7$~kpc corresponds to 
the estimation of the spacing $L$, which was found by \citet{elmegreen1983} 
for their ''H\,{\sc ii} regions'' in Arm~A of NGC~628.

However, the separations discussed above were measured along the arms. And 
for the smallest SFRs and especially for ''the local maxima of brightness'', 
these separations are not distances between adjacent regions. We will return 
to that issue in Conclusion.

\section{Discussion}

Anticorrelation between shock wave signatures and the presence of star 
complexes chains is suspected in spiral arms of a few galaxies. The most 
evident is the M~31 case, where the regular chain of complexes (noted best 
in the {\it GALEX} images) is seen along the NW arm, whereas the stellar age 
gradient across the arm (well established with the Cepheid periods) and 
signatures of the shock wave are observed in another (S4, SW) arm, which 
structure is well described by the classical density wave theory 
\citep{efremov2010}.

The regular wave-like magnetic field is known just in the NW arm 
\citep{beck1989}; therefore the magneto-gravitational instability might lead 
to formation of star complexes chain along this arm segment. Anticorrelation
between the shock wave signature and regular spacing of complexes 
observed in M~31 may be explained with results of \citet{dobbs2008}, 
who concluded that spiral shocks generate an irregular  magnetic field.

\begin{figure*}
\resizebox{1.00\hsize}{!}{\includegraphics[angle=000]{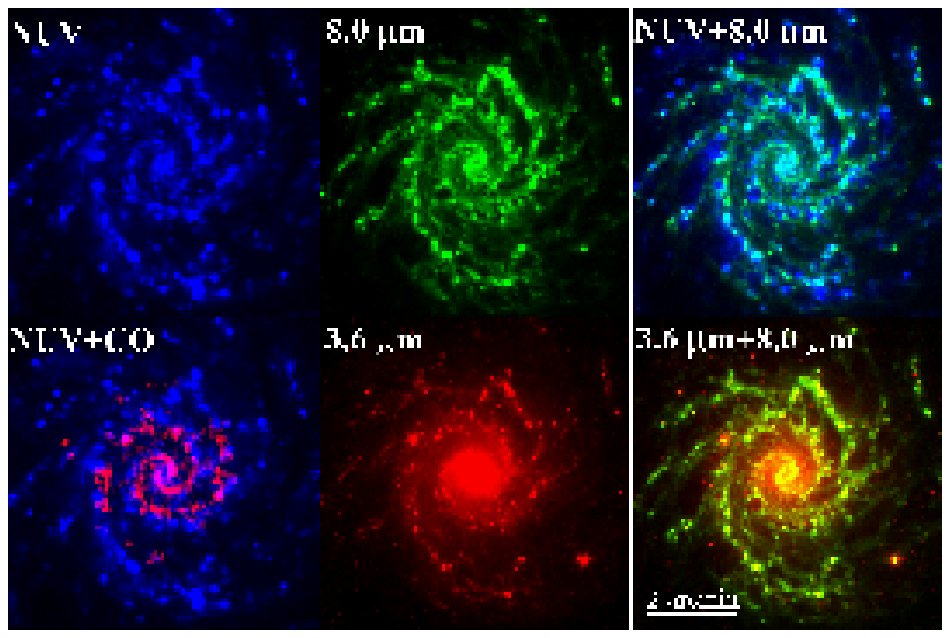}}
\caption{The morphology of the NGC~628 spiral structure. The 
composition of the fits images, taken from the {\sc ned} 
\citep[http://ned.ipac.caltech.edu;][]{dale2009,regan2001,helfer2003}: 
{\it GALEX} $Near~UV$ image (indicator of young stars), 
{\it Spitzer (IRAC)} $8.0~\mu$m image (indicator of dust and PAH molecules), 
{\it Spitzer (IRAC)} $3.6~\mu$m image (indicator of old red giants), 
{\it NRAO} 2.6~mm (CO) image (indicator of H$_2$), and their overlay.
(A colour version of this figure is available in the online version.)
{\bf High resolution jpeg image is available in the source format.}}
\label{figure:efr001f}
\end{figure*}

Amongst nearby galaxies, the most evident (after M~31) case of the 
above-mentioned anticorrelation is NGC~628 (M~74) galaxy. Unfortunately, 
there are no appropriate magnetic field data for the studied part of NGC~628. 
The only data concerning the magnetic field were obtained by 
\citet{heald2009} who detected polarized emission at 18 and 22~cm 
wavelengths from the outer part of the galaxy; their linear beam size 
was $1.9\times0.5$~kpc.

Nevertheless, we can estimate some properties of interstellar medium based 
on the dust distribution in NGC~628. The dust lane along the stellar arm (but 
outside it) can be a signature of the spiral shock wave. In the case of 
NGC~628, we observe the dust lane running in front of inner edge of Arm~B. 
It is clearly seen in the $U$ images of the galaxy (Figs.~\ref{figure:fig1}, 
\ref{figure:nfig5a}), but farther from the center there is no dust lane 
running upstream of the stellar Arm~A inner edge.

To study the disposition of the dust lanes along the spiral arms and 
variation of photometric parameters across the arms, we obtained 
photometric profiles across Arms~A and B. Here, in addition to the 
ultraviolet and optical images, we used $24~\mu$m {\it Spitzer (MIPS)} image 
of the galaxy, taken from the {\sc ned} 
database\footnote{http://ned.ipac.caltech.edu/} \citep{dale2009}, as a good 
tracer of dust location.

Nine positions for the cross-arm photometry were selected, six in Arm~A, 
and three in Arm~B (Fig.~\ref{figure:nfig5a}). Two profiles were obtained 
for each position; the first profile crosses the star formation region, 
and the second profile crosses the arm in the immediate vicinity of a 
SFR. The last set of profiles is marked as 1'--9' (Fig.~\ref{figure:nfig5a}).

The length of across-arm profiles is from 55 to 80~arcsec, the width is equal 
to 3~arcsec for every profile. The fluxes across the P.A. of profile were 
averaged. Obtained profiles 1,~1'--9,~9' of $FUV_0$ surface brightness 
and $(U-B)_0$ colour index, and profiles 1'--9' of $24~\mu$m flux are 
shown in Fig.~\ref{figure:nfig18}. The marks $r=0$~arcsec correspond to 
positions of the logarithmic spiral in Fig.~\ref{figure:nfig5a}.

Photometric profiles in Fig.~\ref{figure:nfig18} clearly illustrate the 
hypothesis of \citet{efremov2010}. Inside Arm~B, the strong dust lane 
is located in the profiles 7,~7' and 8,~8' at $r \approx +10$~arcsec. 
Ultraviolet surface brightness goes down there to 
$\approx 25.5$~mag\,arcsec$^{-2}$, colour index $U-B$ increases to 
$\approx 0.5$~mag, and $24~\mu$m flux reaches a maximum 
(Fig.~\ref{figure:nfig18}). Smoothed dust lane is also observed in 
the profiles 9,~9' at $r \approx +10 - +20$~arcsec. Note that further along 
Arm~B, in its distorted outer part, several large complexes are observed 
(Fig.~\ref{figure:nfig5a}). These complexes are brighter than the regions 
in the studied inner part of Arm~B.

Visually in optical images Arm~A does not show obvious traces of dust 
lane, at least in its outer part (Fig.~\ref{figure:fig1}). However, 
the increase of the colour index $U-B$ at $r = +5$~arcsec in profile 1', 
$r = -11$~arcsec in profiles 2', 5' and 6', and $r = -15$~arcsec in profile 
4' is in accordance with the position of the dust lane inside the stellar 
Arm~A. Maxima of $24~\mu$m flux in profiles 2'--5' are also consistent with 
the middle of the stellar arm (Fig.~\ref{figure:nfig18}).

As Fig.~\ref{figure:nfig18co} demonstrates, the 2.6~mm (CO/H$_2$) 
profiles across spiral arms repeat ultraviolet profiles with a few 
exceptions. In the beginning of the Elmegreens' chain of complexes 
(profiles 2,~2' and 3,~3'), the maximum of the CO flux shifts to the centre 
with respect to the $FUV$ maximum. Fig.~\ref{figure:nfig18co} shows that 
the CO lane is rather located along the inner side of stellar arms (see 
profiles 1,~1', 4,~4', 5,~5', 7,~7').

The overall comparative morphology of spiral arms in NGC~628 is 
demonstrated in Fig.~\ref{figure:efr001f} for the different wave lengths. 
One may see that the CO (2.6~mm, CO/H$_2$) lanes go along the inner sides of 
both stellar arms, but further on along Arm~A, the dust lane goes inside this 
arm and about here the regular spacing of stellar complexes along Arm~A 
appeared (Fig.~\ref{figure:efr001f}). No such a regularity is seen in 
Arm~B, and there is no regular spacing of star complexes as well.

One might think the strong dust lane along the inner side of the stellar 
Arm~B is seen in optics so evidently only due to its bright stellar 
background, but Fig.~\ref{figure:efr001f} demonstrates it is not so -- the 
CO/H$_2$ lane along the inner side of Arm~B is seen very well, whereas it 
absents along the arm outer side. As concerns Arm~A, the age gradient 
(shift between CO and UV arms) is observed to the same distance from the 
center, at which Arm~B becoming quite irregular (Fig.~\ref{figure:efr001f}). 
Comparison of $NUV$, CO, and $8~\mu$m images in 
Fig.~\ref{figure:efr001f} demonstrates, that further from the center in 
Arm~A, the $8~\mu$m and stellar arm central lines coincide.

This is exactly the same situation as observed in the NW segment of 
M~31 spiral arm, where star complexes are seen inside the dust/gas lane. 
There was a suspicion that this M~31 situation might be due to the local 
plane corrugation which might present us the gas lane projected onto the 
stellar arm (whereas the former might be in fact closer to the galaxy center 
in the M~31 plane). Face-on orientation of the M~74 plane makes such an 
explanation untenable.

\section{Conclusions}

Our results confirms the drastic difference in the inner structures 
between the spiral arms of NGC~628, one of which is long and hosts the 
regular chain of star complexes and another does not. 
\citet{elmegreen1983} in the first study of star complexes strings in 
the spiral arms found altogether 22 galaxies with such strings, and only in 
7 of these the chains were noted along both arms; the NGC~628 case seems to 
be typical. The reasons for such a difference are unknown yet. Moreover, it 
is unclear until now why only some $10\%$ of galaxies (all of the grand 
design type) searched by these authors for complexes chains were found to 
host such chains.

This statistics should be enlarged, but one might suggest already that 
indeed the magneto-gravitational instability -- and not just the 
gravitational only -- should be involved in string of complexes formation. 
Magnetic field -- and more so regular field along an arm -- is surely 
more rare phenomenon as comparing with the universal law of gravitation. If 
so, occurrence of the regular chain in the minority of stringed arms only 
may imply we do not know yet some essential properties of intergalactic 
magnetic fields. The cause of drastic difference in inner structures of 
otherwise symmetric arms (in well isolated galaxies!) seems to be amongst 
the greatest unsolved issues in understanding the spiral structures of 
galaxies.

Another our result (which is to be confirmed still) may imply existence of 
a certain minimal distance between young star groups, measured in projection 
onto the arm central line. Many of these groups are too small to be 
called ''a complex'', but most distances are either this minimal one (about 
400~pc, assuming the galaxy distance is 7.2 Mpc -- probably it is larger) or 
twice and four times this minimal distance. This distance is seen 
well within Arm~A between complexes A3 and A6. We cannot exclude the 
possibility that including of associations from Arm~B, which distributed 
there rather chaotically -- and not along the arm -- may lead to erroneous 
(too short) estimate of this ''fundamental'' length.

However, the physical sense of these preferred distances becomes uncertain 
in the central region of the galaxy and this concerns both the arms. 
In this region distances between SFRs, projected onto the central lines 
of both arms cannot be considered as the distance along the arm. This issue 
should be considered in more details. It is quite probable that this central 
region should be excluded from the consideration of the characteristic 
distances between SFRs.

At any rate, we see that both arms of M~74 galaxy are quite different 
(though are symmetrical in relation to the galaxy center). 
The longer Arm~A contains the chain of star complexes, whereas the shorter
Arm~B hosts the (irregular) chains of (rather small) H\,{\sc ii} regions as
well as dust lanes, one quite strong along the inner side of the stellar 
arm and another lane, much lesser expressed along the outer side. 
Implications of the later phenomena should be studied yet.

The last unsolved issue we want to mention here is missing of a 
complex (i.e. the double distance between adjacent complexes) in a string, 
which was noted already in \citet{elmegreen1983}. They found that some of 
their ''H\,{\sc ii} regions'' have double mutual distance as comparing 
with the average expected between two ones along a given arm. Similarly, 
later \citet{efremov2010} found that some distances between regularly spaced 
H\,{\sc i} superclouds along the Carina arm of the Galaxy are twice larger 
than all others. These facts are hardly possible to explain in other way 
apart from the gravitational or, rather, magneto-gravitational instability, 
developing along the arm in the gas tube to form superclouds, the parent 
bodies for star complexes. Unfortunately, the HST data are not available 
for the outer part of Arm~A, where the regular chain of complexes is observed.

\section*{Acknowledgments}

We are grateful to the referee for his/her constructive comments. 
The authors would like to thank E.~V.~Shimanovskaya (SAI MSU) for help with 
editing this paper. The authors acknowledge the usage of the HyperLeda data 
base (http://leda.univ-lyon1.fr), Barbara A. Miculski archive for space 
telescopes (http://galex.stsci.edu) and the NASA/IPAC Extragalactic Database 
(http://ned.ipac.caltech.edu). This study was supported in part by 
the Russian Foundation for Basic Research (project no. 12--02--00827).

\end{document}